\documentclass[prd,aps,10pt,twocolumn,floatfix,nofootinbib]{revtex4-1}
\usepackage{lipsum}
\usepackage{textcomp}
\usepackage{amsmath}
\usepackage{amsfonts}
\usepackage{amssymb}
\usepackage{graphicx}
\usepackage{hyperref}
\usepackage{subfigure}
\usepackage[usenames, dvipsnames]{color}
\usepackage[utf8]{inputenc}
\providecommand{\abs}[1]{\left\lvert#1\right\rvert} % modulo
\providecommand{\norm}[1]{\left\lVert#1\right\rVert} % norma
\newtheorem{theorem}{Theorem}[section]

\newtheorem{definition}{Definition}[section]

\begin{document}
\title{On the ill posedness of Force-Free Electrodynamics in Euler Potentials}
\author{Oscar A. Reula}\email{reula@famaf.unc.edu.ar}
\author{Marcelo E. Rubio}\email{merubio@famaf.unc.edu.ar}
%\homepage{http://legauss.blogspot.com}
\affiliation{Facultad de Matem\'atica,
Astronom\'{\i}a, F\'{\i}sica y Computaci\'on, Universidad Nacional de C\'ordoba,\\
Instituto de F\'{\i}sica Enrique Gaviola, CONICET.\\
Ciudad Universitaria, (5000) C\'ordoba, Argentina}

\date{\today}

\begin{abstract}
We prove that the initial value problem for Force-free Electrodynamics in Euler variables is not well posed. 
We establish this result by showing that a well-posedness criterion provided by Kreiss fails to hold for this theory, and using a theorem provided by  Strang. To show the nature of the problem we display a particular bounded (in Sobolev norms) sequence of initial data for the Force-free equations such that at any given time as close to zero as one wishes, the corresponding evolution sequence is not bounded. Thus, the Force-free evolution is non continuous in that norm with respect to the initial data. We furthermore prove that this problem is also ill-posed in the Leray-Ohya sense.
\end{abstract}

\maketitle

%\tableofcontents

%%%%%%%%%%%%%%%%%%%%%%%%%%%%%%%%%%%%%%%%%%%%%%%%

\section{Introduction}

%%%%%%%%%%%%%%%%%%%%%%%%%%%%%%%%%%%%%%%%%%%%%%%%

It is well known that in the neighborhood of a pulsar or a black hole, the presence of strongly magnetic fields gives rise to the generation of a very diluted plasma. In that region, the electromagnetic field dominates over the matter constituting those plasma, and the resulting uncoupled dynamics is commonly known as \textit{Force-free Electrodynamics} (FFE).

A complete theory of FFE has been developed in several works. A recent review is presented in \cite{Gralla14}, in which many theoretical aspects of the corresponding dynamics are generalized from a relativistic perspective, focusing on intrinsic geometrical properties of the theory, and providing further results. In a previous work series by Uchida, \cite{UchidaI97, UchidaII97}, a covariant formulation of FFE without spacetime symmetries assumptions is developed, using \textit{Euler Potentials} as evolution variables. Euler Potentials are scalar functions that were introduced by Stern \cite{Stern70} in the early 70's from a non covariant formulation. This formulation often appears in numerical simulations, see for example \cite{Brandenburg11, Zaharia08}.

In \cite{Komissarov12}, Komissarov focused on the hyperbolicity of general degenerated force-free electromagnetic theories from a non covariant viewpoint, in which the evolution equations were presented in the form of conservation laws. Subsequently, Pfeiffer \cite{Pfeiffer13} modified Komissarov's equations to obtain a symmetric-hyperbolic evolution system. Very recently, in \cite{Carrasco16}, Geroch's geometric formalism of symmetric-hyperbolic systems was used to introduce a covariant hyperbolization of FFE in Maxwell variables, and a detailed analysis of the characteristic structure of the evolution system as well as the resulting causal cone structure was performed. In that work, the authors used the same techniques successfully employed in \cite{Abalos15} for  non-linear generalizations of Maxwell's theory of Electromagnetism, in which FFE is not included there (see \cite{Carrasco16} for more details).

In this work we study the problem of hyperbolicity for the version of Force-free Electrodynamics in Euler Potentials (FFEEP). We prove that FFEEP is not strongly hyperbolic, that is, the associated Cauchy problem is not well posed. In particular, it is not possible to find a hyperbolizer (that is, a symmetric, non degenerate and positive definite bilinear form) for the evolution equations like the one found in \cite{Carrasco16}. To do so, we make use of an algebraic criterion introduced by Kreiss \cite{Kreiss89} that provides necessary and sufficient conditions for a square first order system to be well-posed. From a detailed analysis of the characteristic structure of the dynamic equations, we see that there is not a complete set of eigenvectors of the corresponding algebraic problem (characteristic equations), allowing us to prove that Force-free evolution is generally non continuous with respect to the initial data, in any Sobolev norm. We illustrate this feature by constructing a bounded sequence of initial data such that, for any time as close to zero as desired, the corresponding evolution sequence is not bounded. Finally, we study the problem of well-posedness of FFEEP in the sense of Leray and Ohya, and we prove that this system is also ill-posed in that sense.

\subsection{Outline, units and conventions}

This paper is organized as follows. In section \ref{II} we present a brief review about Force-free Electrodynamics as a degenerate theory, and several geometric and algebraic properties that arise from this fact are discussed. Furthermore, we introduce Euler Potentials and the notion of flux surfaces in spacetime. In section \ref{III} we present the evolution system that will be treated along this work. In particular we discuss gauge freedom, hyperbolicity and wave-set structure of the dynamic equations. Section \ref{IV} is devoted into studying the ill posedness of this version of FFE. We first show the failure of a necessary and sufficient criterion for squared first order systems to be well posed, then we perform a 3+1 decomposition of the evolution equations that will be used later on, and we prove, with an explicit example, lack of continuity (in Sobolev norms) of the evolution with respect to the initial data. Finally, we study the hyperbolicity of FFEEP in the sense of Leray and Ohya. Appendix \ref{app1} is dedicated to review the main ideas about hyperbolicity in the Leray-Ohya sense, and to fix notation and definitions we shall adopt.

The signature convention of the spacetime metric we will use along this work is $(-,+,+,+)$, and we will take units such that $c = G = 1$, where $c$ is the speed of light in vacuum and $G$ Newton's universal constant of gravitation. For Maxwell's equations, we will adopt gaussian units.

%%%%%%%%%%%%%%%%%%%%%%%%%%%%%%%%%%%%%%%%%%%%%%%%

\section{Preliminaries}
\label{II}
%\setcounter{equation}{0}

%%%%%%%%%%%%%%%%%%%%%%%%%%%%%%%%%%%%%%%%%%%%%%%%

\subsection{Force-free Electrodynamics}

%%%%%%%%%%%%%%%%%%%%%%%%%%%%%%%%%%%%%%%%%%%%%%%%

Force-free Electrodynamics is a non linear version of Maxwell's equations imposing the \textit{force-free approximation}. Recall that Maxwell's equations are given by
\begin{equation}\label{eq:max_eq's}
{\left\{
{\begin{array}{rcl}
\nabla_a F^{ab} & = & - 4\pi J^{b}\;;\\
\nabla_{[a} F_{bc]} & = & 0 \;,
\end{array}}
\right.}
\end{equation}
where $F_{ab}$ is an antisymmetric $(0,2)$ smooth tensor field over a background spacetime $(\mathcal{M},\;g_{ab})$, and $J^a$ the electromagnetic 4--current, which is conserved by virtue of the antisymmetry of $F_{ab}$. Associated to the theory there is an electromagnetic energy-momentum tensor, namely
\begin{equation} \label{eq:en-mom-tensor}
T_{ab} = \frac{1}{4\pi}\left(F_{ac} F_{b}{}^c - \frac{1}{4} g_{ab} F^{cd} F_{cd}\right).
\end{equation}
In the presence of matter this tensor is in general not conserved, by virtue of (\ref{eq:max_eq's}). Indeed, 
\begin{equation}\label{eq:conservation}
\nabla_b T^{ab} = - F^{a}{}_b J^{b},
\end{equation}
and it implies that locally, the energy-momentum loss is just the work exerted by the electric force $\rho E$ times the velocity of the
particles.

In the context of \textit{relativistic magnetohydrodynamics}, the total energy-momentum tensor contains the contributions both of matter and electromagnetic field. The \textit{force-free approximation} consists in neglecting the matter contribution in situations where it is orders of magnitude smaller than the  electromagnetic contribution\footnote{In \cite{Komissarov12}, Force-free Electrodynamics is defined as the zeroth order system of relativistic magnetohydrodynamics under matter perturbations.}. Thus, by total conservation of energy-momentum tensor, we get
\begin{equation}\label{eq:FF_aprox}
F_{ab} J^{b} = 0,
\end{equation}
from which it follows that $F_{ab}$ is a \textit{degenerate} 2--form. This property implies that there is a frame (defined by $J^a$) in which the electric field vanishes. Such approximation is relevant in general electromagnetic systems which are \textit{magnetically dominated}\footnote{In general, it is not expected that the condition of magnetic dominance is preserved during evolution.}, that is, when the efect of electric currents are neglected in comparison those of magnetic presures, then the system becomes a plasma whose only effect is, due to its conductivity, to make the electric field much smaller than the magnetic field.

Thus, Maxwell's field can be evolved independently of matter degrees of freedom, and a complete set of equations for the electromagnetic field is obtained by adding to system (\ref{eq:max_eq's}) the force-free condition (\ref{eq:FF_aprox}). Indeed, contracting the first equation in (\ref{eq:max_eq's}) with $F_{bc}$ we get
\begin{equation}\label{eq:FFsystem}
{\left\{
{\begin{array}{rcl}
F_{bc}\nabla_a F^{ab} & = & 0\;;\\
\nabla_{[a} F_{bc]} & = & 0\;.
\end{array}}
\right.}
\end{equation}
Notice that solutions of vacuum Maxwell's equations are trivially solutions of (\ref{eq:FFsystem}), but in general there are more solutions and they behave quite differently form Maxwell's. 

From a pure algebraic viewpoint, condition (\ref{eq:FF_aprox}) means that $\det(\textbf{F}) = 0$, or which is the same,
\begin{equation}\label{eq:*FF=0}
{}^{*}F^{ab} F_{ab} = 0,
\end{equation}
where ${}^{*}F^{ab}$ is the action on $F^{ab}$ of the Hodge operator $*$, given by
\begin{equation}\label{eq:dual_F}
{}^{*}F^{ab} = \frac{1}{2}\varepsilon^{abcd}F_{cd},
\end{equation}
and $\varepsilon_{abcd}$ is the volume element compatible with $g_{ab}$. This condition leads to interesting geometric properties of force-free systems, that will be developed in a particular case in the next section. Straightforwardly, the antisymmetry and degeneracy of Maxwell's tensor implies that the kernel of $F^a{}_b$ is a two-dimensional vector space and thus there exist two linearly independent vectors $\{u^a,\;v^a\}$ such that
\begin{equation} \label{eq:u,v_li_ker(F)}
F^a{}_b\; u^b = F^a{}_b\; v^b = 0. 
\end{equation}

Assuming some regularity for $F_{ab}$, one can show that both vector fields that satisfy property (\ref{eq:u,v_li_ker(F)}) are \textit{integrable}; that is, they smoothly generate a 2--dimensional surface, known as the \textit{flux surface} or \textit{field sheet}. This \textit{integrability condition}\footnote{More explicitly, the kernel of $F_{ab}$ is \textit{integrable} in the above sense if it is tangent to two--dimensional submanifolds of $(\mathcal{M},\; g_{ab})$.} follows from the fact that $\nabla_{[a} F_{bc]} = 0$. The \textit{magnetic field line} measured by any observer $t^a$ corresponds to the intersection between the flux surface and the observer's hypersurface $\{t = \mbox{const}\}$. Since $F_{ab} F^{ab} > 0$, such a flux surface is temporal and so it is possible to interpret it as the \textit{world sheet} of the initial magnetic field line during evolution. In particular, equation (\ref{eq:FF_aprox}) implies that $J^a$ is \textit{tangent} to it.

Another relevant aspect of force-free systems is a trivial consequence of equation (\ref{eq:FF_aprox}). Indeed, the equality $F_{[ab} F_{cd]} J^d = 0$ holds, and since $F_{[ab} F_{cd]}$ is a 4--form over a 4--dimensional manifold, it must be proportional to the associated volume element, which is a \textit{non} degenerate tensor. Thus, there must be $F_{[ab} F_{cd]} = 0$, which implies\footnote{See \cite{Penrose84}, prop. 3.5.35.} that $F_{ab}$ is \textit{simple}, i.e., there exist two covector fields $\ell^1_a$, $\ell^2_a$ such that, locally,
\begin{equation}\label{eq:2ab}
F_{ab} = 2 \ell^1_{[a} \ell^2_{b]}.
\end{equation}

It is clear that $\ell^1$ and $\ell^2$ are linearly independent, and for magnetically dominated systems, both are space-like vector fields. Recalling now property (\ref{eq:u,v_li_ker(F)}) we see that $\ell^1$ and $\ell^2$ are \textit{orthogonal} to the flux surfaces.

%%%%%%%%%%%%%%%%%%%%%%%%%%%%%%%%%%%%%%%%%%%%%%%%

\subsection{Euler Potentials}
\label{euler-pot-section}
%%%%%%%%%%%%%%%%%%%%%%%%%%%%%%%%%%%%%%%%%%%%%%%%

Integrability condition of the kernel of $F^a{}_b$ also implies the local existence of two \textit{scalar fields}, say $\{\phi_1, \;\phi_2\}$ such that flux surfaces are described\footnote{See also \cite{Carter79}.} by the intersection of the level set of these to functions, $\phi_1 = \mbox{const.}$ and $\phi_2 = \mbox{const.}$. These functions are commonly known as \textit{Euler Potentials} and share several interesting properties. By the local expression (\ref{eq:2ab}), it must be 
\begin{equation}\label{eq:max_poteu}
F_{ab} = \nabla_a\phi_1 \nabla_b\phi_2 - \nabla_b\phi_1 \nabla_a\phi_2,
\end{equation}
and the vector potential in Euler variables reads 
\begin{equation}\label{eq:max-poteuvec}
A_a = \frac{1}{2}\left(\phi_1 \nabla_a\phi_2 - \phi_2 \nabla_a\phi_1\right). 
\end{equation}

For $i=1,\;2$, let us denote the normal vector fields of the flux surfaces as
\begin{equation}
\ell^a{}_i := g^{ab} \nabla_b \phi_i.
\end{equation}

At this point, it is convenient to introduce certain internal structure for ease of notation. For $i=1,2$, let $\varepsilon^{ij}$ be an antisymmetric symbol such that $\varepsilon^{12} = 1$, and assume that there is an inverse, $\varepsilon_{ij}$, with the following property:
\begin{equation} \label{eq:epsilon-delta}
\varepsilon^{ij} \varepsilon_{jk} = -\;\delta^{i}{}_k.
\end{equation}
Here, $\delta^{i}{}_j$ is the identity map, that is $\delta^{i}{}_j A^{j} = A^{i}$. For a given $f_i$, let us denote $f^{i}:=\varepsilon^{ij}f_j$. Inversely, and by consistence with (\ref{eq:epsilon-delta}), $f_i = -\varepsilon_{ij} f^j$. Such an internal structure allows one to express the fields (\ref{eq:max_poteu}) and (\ref{eq:max-poteuvec}) in a more convenient way. Indeed,
\begin{equation} \label{eq:max_pot_sympl}
F_{ab} = \varepsilon^{ij} \nabla_a\phi_i \nabla_b\phi_j\;,
\end{equation}
and
\begin{equation}
A_a = \frac{1}{2} \varepsilon^{ij}\phi_i \nabla_a\phi_j.
\end{equation}

Notice that there is certain freedom in the choice of Euler potentials such that $F_{ab}$ remains invariant. In effect, if one considers the following general transformation:
\begin{equation} \label{eq:gauge_transf}
\phi_i \; \mapsto \; \tilde{\phi_j} = \tilde{\phi_j}(\phi_i),
\end{equation}
the gradients change as
\begin{equation}\label{eq:grad_transf}
\nabla_a \tilde{\phi}_j = \frac{\partial \tilde{\phi}_j}{\partial \phi_i} \nabla_a \phi_i := \chi^{i}{}_{j} \nabla_a \phi_i,
\end{equation}
where we have denoted
\begin{equation}
\chi^i{}_j := \frac{\partial \tilde{\phi}_j}{\partial\phi_i}.
\end{equation}
On the other hand, Maxwell's tensor transforms as
\begin{align} \label{eq:max_transf}
\tilde{F}_{ab} & = \varepsilon^{ij} \nabla_a \tilde{\phi_i} \nabla_b \tilde{\phi_j} \nonumber \\ 
& = \varepsilon^{ij} \chi^{k}{}_i \chi^{\ell}{}_j \nabla_a \phi_k \nabla_b \phi_{\ell} \nonumber \\ 
& = \det\left(\chi\right) F_{ab}, 
\end{align} 
where in the last step we used the fact that any (2,0) antisymmetric symbol in the internal space is proportional to $\varepsilon^{ij}$. The proportionality factor is exactly the determinant of $\chi$. Thus, only transformations like (\ref{eq:gauge_transf}) with $\det\left(\chi\right) = 1$ leave $F_{ab}$ invariant. These are transformations that belong to the $SL(2,\mathbb{R})$ group. 

This condition has also a relevant geometric significance. Indeed, it is straightforward to see that
\begin{equation} \label{eq:det-area}
\det\left(\chi\right) = \abs{\frac{\partial (\tilde{\phi}_1,\tilde{\phi}_2)}{\partial\left(\phi_1, \phi_2\right)}},
\end{equation}
so the invariance condition is equivalent to require that the jacobian of transformation (\ref{eq:gauge_transf}) is the unity. This is a necessary and sufficient condition for the area element to not change under a coordinate change.

Finally, we will find it useful to introduce the bilinear form given by
\begin{equation}
G_{ij} := \ell_i \cdot \ell_j = g^{ab} \nabla_a \phi_i \nabla_b \phi_j.
\end{equation}
By construction $G_{ij}$ is symmetric, and satisfies
\begin{align} \label{eq:detG}
\det\left(G\right) & = \frac{1}{2}\varepsilon^{ij} \varepsilon^{k m} G_{ik} G_{j m} \nonumber \\
& = \frac{1}{2} \left(\varepsilon^{ij}\ell^a_i \ell^b_j\right) \left(\varepsilon^{km} g_{ac}\ell^c{}_k g_{bd} \ell^d{}_{m}\right) \nonumber \\
& = \frac{F}{2},
\end{align}
where $F := F^{ab} F_{ab}$ is a positive function for magnetically dominated systems. Thus, $\det\left(G\right) \neq 0$ in general and $G_{ij}$ is invertible. Namely, there exists a symmetric quantity $\tilde{G}^{ij}$ such that $\tilde{G}^{ij} G_{jk} = \delta^{i}{}_k$. On the other hand, notice that there is another way to construct a symmetric symbol of rank (2,0) from $G_{ij}$, namely, $G^{ij}:=\varepsilon^{ik} \varepsilon^{j\ell} G_{k \ell}$. A straightforward calculation shows that
\begin{equation} \label{eq:tildeG=G}
\tilde{G}^{ij} = \frac{2}{F} G^{ij},
\end{equation}
so both quantities contain the same information up to a scale factor.

%%%%%%%%%%%%%%%%%%%%%%%%%%%%%%%%%%%%%%%%%%%%%%%%

\subsection{Well-posedness}
\label{s:kreiss_theorem}

%%%%%%%%%%%%%%%%%%%%%%%%%%%%%%%%%%%%%%%%%%%%%%%%

This section contains a brief review of the main ideas about well posed systems in Physics. In particular, we introduce the notions of \textit{hyperbolicity}, \textit{strong}, \textit{symmetric}, and \textit{weak} first order systems, as well as the basic notions of well-posed and ill-posed systems. 
We shall follow the theory and definitions given in \cite{Kreiss89, Kreiss-Ortiz01, Reula98, Reula04, Geroch96, Kreiss70} and provide some essential definitions for general quasi-linear first order systems.

One of the fundamental questions that arise in understanding the evolution of dynamical systems in physics is their hyperbolicity.
This concept captures some aspects that should hold even in the most fundamental scenarios, and its understanding leads one to answer questions about: uniqueness of solutions for a given initial data, preservation of the asymptotic decay of the solution with respect to that of their initial data, and estimates about time of existence of the solutions, among others. All of these aspects are related to the \textit{continuity} of the map that goes from the set of initial data to the set of solutions.

To start, let us consider the linear constant-coefficient problem given by
\begin{equation}\label{eq:kreiss_systems-one}
{\left\{
{\begin{array}{rcl}
\partial_t u & = & A^i \partial_i u =: P(\partial)\; u\;; \\
u(x,0) & = & f(x)\;,
\end{array}}
\right.}
\end{equation}
where $u = u(x,t) \in \mathbb{C}^{s}$, $x = (x_1,\cdots,x_n)$ are space coordinates, and $A^i$ is a $s\times s$ constant complex matrix valued vector in $\mathbb{R}^n$. The main purpose that drives us to deal with these problems is to give necessary and sufficient conditions for the associated Cauchy problem to be \textit{well posed}, that is, under what conditions there exists a unique solution of (\ref{eq:kreiss_systems-one}) that depends continuously on the initial data. It turns out that under certain circumstances those conditions are also valid for more general systems, namely those where the matrices $A^i$ are smooth functions of $u$, like (\ref{eq:kreiss_systems}). It follows also that if the constant-coefficient equation system is ill-posed for some values of $A^i$, any quasi-linear system having in a point those values of $A^i(u)$ will be also ill-posed, as we shall assert later on.

Let us now restrict the possible initial data of (\ref{eq:kreiss_systems-one}) into complex functions $f(x)$ of the form
\begin{equation}\label{eq:plane-waves}
f(x) = \frac{1}{(2\pi)^{n/2}} \int_{\mathbb{R}^n}{e^{i k \cdot x} \hat{f}(k)\; d^n k},
\end{equation}
where $\hat{f}(k)$ is of compact support. The unique smooth solution of (\ref{eq:kreiss_systems-one})-(\ref{eq:plane-waves}) is
\begin{equation}\label{eq:sol-kreiss}
u(x,t) = \frac{1}{(2\pi)^{n/2}} \int_{\mathbb{R}^n}{e^{i k \cdot x} e^{P(i k) t} \hat{f}(k)\; d^n k},
\end{equation}
where $P(i k)$, which is formally obtained by substitution of $i k_j$ for $\partial / \partial x_j$, is called the \textit{symbol}\footnote{We refer the reader to \cite{Taylor91} in which pseudo-differential analysis is discussed in detail.} of $P(\partial)$.

\begin{definition}
\label{well-posed}
System (\ref{eq:kreiss_systems-one}) is called \textit{well posed} if there exists a unique solution in a neighborhood of $t=0$, and that solution depends continuously on the initial data; that is, there exists a norm $\norm{\;\cdot\;}$ and two constants $C$, $\alpha$ such that
for all initial data like (\ref{eq:plane-waves}) and $t > 0$,
\begin{equation}
\norm{u(x,t)} \leq C e^{\alpha t} \norm{f(x)}.
\end{equation}
\end{definition}

In order to characterize well-posed systems like (\ref{eq:kreiss_systems-one}), it suffices to give algebraic conditions on the principal part of the equations, such that the corresponding Cauchy problem is well-posed. There are several notions of hyperbolicity. We shall review here such notions that will be relevant along this work.

\begin{definition}
\label{str-hyp}
System (\ref{eq:kreiss_systems-one}) is called \textit{strongly hyperbolic} if for any covector $k_c$, the matrix $\mathbb{A} := A^i k_i$ has only purely real eigenvalues and is diagonalizable.
\end{definition}

Due to the fact that any complex matrix $A$ is diagonalizable with only real eigenvalues if and only if there exists a \textit{symmetrizer} $H$, that is, a positive definite bilinear form, such that $HA$ is symmetric, it follows that (\ref{eq:kreiss_systems-one}) is strongly hyperbolic if and only if for each $k_a$ there is a matrix $H(k)$ such that $H(k)\mathbb{A}$ is symmetric.

\begin{definition}
\label{sym-hyp}
System (\ref{eq:kreiss_systems-one}) is \textit{symmetric hyperbolic} if it is possible to find a \textit{common symmetrizer} $H$ for all possible $k_c$.
\end{definition}

Notice that the notion of symmetric hyperbolicity is a sufficient but not necessary condition to guarantee well posedness. Strong hyperbolicity, however, ensures well-posedness of the initial value problem in the sense of definition \ref{str-hyp}, that is, once a particular Sobolev norm has been chosen. 

A set of important and clarifying results about well posedness for \textit{constant-coefficient} first order systems is provided by Kreiss in \cite{Kreiss89}. These results reduce the problem of well-posedness into a pure algebraic issue.

\begin{theorem}
A system like (\ref{eq:kreiss_systems-one}) is \textit{well posed} if and only if there exist constants $C$ and $\alpha$ such that for all $t > 0$,
\begin{equation}
\abs{e^{P(i k)t}} \leq C e^{\alpha t},
\end{equation} 
for all $k \in \mathbb{R}^n$, where $\abs{\;\cdot\;}$ is the usual matrix norm.
\end{theorem}

\begin{theorem}
\label{kreiss}
Let \textbf{F} denote a set of matrices $A\in\mathbb{C}^{n\times n}$. The following conditions are equivalent:
\begin{itemize}
\item There is a constant $K_1$ with $|e^{At}|\leq K_1$ for all $A\in$ \textbf{F} and all $t \geq 0$.

\item For all $A\in$ \textbf{F} and all $s\in\mathbb{C}$ with Re$(s)>0$ the matrix $A-sI$ is non singular, and there is a constant $K_2$ such that
\[
|(A-sI)^{-1}|\leq\frac{K_2}{\mbox{Re}(s)}, \quad A\in\mbox{\textbf{F}}, \quad \mbox{Re}(s)>0.
\]
(This condition is known as the \textbf{Resolvent Condition}).

\item There is a positive constant $c$ with the following property: for each $A\in$ \textbf{F}, there exists a Hermitian matrix $H=H(A)\in\mathbb{C}^{n\times n}$ with
\[
c^{-1}I_{n\times n} \leq H \leq c\;I_{n\times n} \qquad \mbox{\textit{and}} \qquad HA + AH^{*} \leq 0.
\]
\end{itemize}
\end{theorem}

If the system is such that the matrix $\mathbb{A}=A^ik_i$ previously introduced in Definition \ref{str-hyp} has imaginary eigenvalues but their eigenvectors do not form a base (this is the case if $\mathbb{A}$ is \textit{not} diagonalizable), then the system is called \textit{weakly hyperbolic}. In some particular cases one can show that the system is \textit{weakly well posed} in the sense of Kreiss \cite{Kreiss89}; namely that they satisfy an inequality of the following form:
\begin{equation}\label{eq:w-hyp-ineq}
\abs{e^{P(ik)t}}\leq \beta\left[1 + \left(|\vec{k}|t\right)^{\gamma}\right] e^{\alpha t},
\end{equation}
for some real constants $\alpha$, $\beta$, $\gamma$ and $t \geq 0$. These type of systems are characterized by the appearance of terms that grow polynomially in $|\vec{k}|t$ in the matrix exponential, $e^{P(ik)t}$, and so cannot be bounded independently of $|\vec{k}|$. The above inequality means that the solution is a continuous function of the initial data but in different topologies, i.e., Sobolev spaces of different orders. This is not a pleasurable situation for it means that every time we restart an iteration (say for solving the system via approximations), we loose derivatives, becoming the solution less and less smooth. In any case, estimate (\ref{eq:w-hyp-ineq}) is very rare to encounter; it often happens that generic lower order perturbations (which might arise for instance from treating variable-coefficient systems and making higher order energies) destroy it. A clarifying example can be found in \cite{Kreiss-Ortiz01}, after Definition 2, in which the corresponding system is well posed but in the weak sense (\ref{eq:w-hyp-ineq}), lower order terms that can be added to it, causing an exponential growth (in frequency) of the solution.

In this work we will see that the constant-coefficient rendition of Force-free equations with Euler Potentials is only \textit{weakly hyperbolic}, and we shall refer these sort of systems as \textit{ill-posed} systems.

As previously asserted, all the results provided above can be generalized to general \textit{quasi-linear systems}, i.e, systems like
\begin{equation}\label{eq:kreiss_systems}
{\left\{
{\begin{array}{rcl}
\partial_t u^{\alpha} & = & A^{\alpha c}{}_{\beta}(t,x,u) \partial_c u^{\beta} + B^{\alpha}(t,x,u)\;; \\
u^{\alpha}|_o & = & f^{\alpha}\;,
\end{array}}
\right.}
\end{equation}
where $u^{\alpha} = u^{\alpha}(t,x)$ are unknown arbitrary tensor fields, and $A^{\alpha i}{}_{\beta}$ and $B^{\alpha}$ depend smoothly on their arguments. While the behavior of solutions of these type of problems is not yet fully understood, it is possible to use the constant-coefficient theorems to prove well posedness for small time intervals with a correspondingly generalized definition of well-posedness. 

\begin{definition}
\label{ql-wp}
For $0< T_o < \infty$, let $u_o (t, x)$, $t\in [0, T_o)$ be a smooth solution of a
quasi–linear evolution system like (\ref{eq:kreiss_systems}). We shall say the system is well-posed at the
solution $u_o$ and with respect to a norm $\norm{\;\cdot\;}$ if given any $\delta > 0$ there exists $\varepsilon > 0$ such that for any smooth initial data $f (x)$ with $\norm{f - f_o}< \varepsilon$, where $f_o(x) := u_o (0, x)$, there exists a smooth solution $u(t, x)$ defined in a strip $0 \leq t < T$, that satisfies $\abs{u(t,\cdot) - u_o(t,\cdot)} < \delta$ when $\abs{T-T_o}<\delta$.
\end{definition}

To discuss the hyperbolicity of general quasi-linear systems, modifications of some properties stated in the constant-coefficient case must be made. Indeed, there is the following

\begin{definition}
System (\ref{eq:kreiss_systems}) is \textit{strongly hyperbolic} if there exists a \textit{symmetrizer}, that is, a symmetric, positive definite matrix $H_{\alpha \beta} = H_{\alpha \beta} (t,x,u,k)$, depending smoothly on its arguments, such that $h_{\alpha \beta}:=H_{\alpha \gamma} A^{\gamma c}{}_{\beta} k_c$ is also symmetric for all one-forms $k_c$.
\end{definition}

Again, via the smoothness required both on the coefficients in (\ref{eq:kreiss_systems}) as in $H_{\alpha \beta}$, strong hyperbolicity holds if and only if the \textit{principal part} $ A^{\alpha c}{}_{\beta}(t,x,u) k_c$ has a complete set of eigenvectors with purely real eigenvalues. Nevertheless, in order to prove that a quasi-linear system is \textit{not} well-posed, a simple and useful result provided by Strang in \cite{Strang66} can be employed. In that work, the author deals with higher order systems of the form
\begin{equation} \label{eq:strang-syst}
\partial_t u = \sum_{\abs{\alpha}\leq m}{A_{\alpha}(x)D^{\alpha}u},
\end{equation}
where $x=(x_1,\cdots,x_n)\in\mathbb{R}^n$, $u=u(t,x)\in\mathbb{C}^s$, $\alpha := (\alpha_1,\cdots,\alpha_n)\in\mathbb{N}_o^{n}$ and
\[
D^{\alpha} := \frac{\partial^{\abs{\alpha}}}{\partial x_1^{\alpha_1}\cdots \partial x_n^{\alpha_n}}, \qquad \abs{\alpha} := \alpha_1 + \cdots + \alpha_n.
\]
Strang asserts that if a system like (\ref{eq:strang-syst}) is well posed in $L^2$ norm, then their principal parts, with the coefficients evaluated at any point and any solution close to the initial data, must be also well-posed in the sense of definition \ref{well-posed} (see \cite{Strang66} for details). The relevance of this theorem is the fact that it shows that the problem of well-posedness is a micro-local (or \textit{high frequency} phenomena, as we shall see) issue. Since for quasi-linear systems the principal part coincides with the linearization at an arbitrary point, if we take the linearization of the system around a constant solution and we prove that the resulting system is not well-posed, then the full system will be not well-posed. We will use this technique to show that force-free system written in Euler variables is in general non continuous with respect to the initial data.

%%%%%%%%%%%%%%%%%%%%%%%%%%%%%%%%%%%%%%%%%%%%%%%%

\section{The system}
\label{III}
%\setcounter{equation}{0}

%%%%%%%%%%%%%%%%%%%%%%%%%%%%%%%%%%%%%%%%%%%%%%%%

\subsection{The equations}

%%%%%%%%%%%%%%%%%%%%%%%%%%%%%%%%%%%%%%%%%%%%%%%%

Rewriting system (\ref{eq:FFsystem}) by expressing Maxwell's tensor in terms of Euler potentials $\phi_i$, $i=1,2$, we get the following equation system:
\begin{equation}\label{eq:ffe-eq's-full}
\varepsilon^{ij} \nabla_a \phi_k \nabla_c (\nabla^a \phi_i \nabla^c \phi_j) = 0, \quad k = 1,2,
\end{equation}
where $\varepsilon^{ij}$ and all the internal structure we shall use was previously introduced (see section \ref{euler-pot-section}).

It is straightforward to see that (\ref{eq:ffe-eq's-full}) is invariant under unitary gauge transformations like (\ref{eq:gauge_transf}). Moreover, once the gauge choice is made in one space-like hypersurface, then \textit{it will remain fixed for all time}. This important property is a direct consequence of the feature that Euler potentials are \textit{constant} along each magnetic world sheet. In particular, they are constant at the intersection of the flux surface and any Cauchy hypersurface $\Sigma$. Thus, once the initial data of (\ref{eq:ffe-eq's-full}) is given, and so the gauge choice is made at each point of $\Sigma$, the gauge transformation will remain constant during evolution.

Indeed, if $\phi_i\mapsto\tilde{\phi}_i$ is any transformation like (\ref{eq:gauge_transf})-(\ref{eq:grad_transf}), then locally we have
\begin{align}\label{eq:cuenta}
0 & = \nabla_{[a}\nabla_{b]}\tilde{\phi}_k \nonumber \\
& = \left(\nabla_{[a} \chi_{|k|}{}^j\right) \ell_{|j| b]} + \chi_{k}{}^j \nabla_{[a} \ell_{|j| b]} \nonumber \\
& = \left(\nabla_{[a} \chi_{|k|}{}^j\right) \ell_{|j| b]}.
\end{align}

Taking now a vector field $t^a$ in the orthogonal complement of $\{\ell^a{}_1,\; \ell^a{}_2\}$, i.e. tangent to flux surfaces, and contracting (\ref{eq:cuenta}) with $t^a \ell^b{}_i$, we get

\begin{equation}
(t^a \nabla_{a} \chi_k{}^j)\; G_{ij} = 0,
\end{equation}
or equivalently, since $G_{ij}$ is invertible (see equation (\ref{eq:tildeG=G})),
\begin{equation} \label{eq:conserv_gauge}
t^a \nabla_a \chi_k{}^j = 0.
\end{equation}

Thus, the functions $\chi_i{}^j$ are constant along the field surfaces $\phi_i=\mbox{const}$, as expected. Moreover, once the gauge choice is done in one space-like hypersurface, then it will remain fixed for all time.

%%%%%%%%%%%%%%%%%%%%%%%%%%%%%%%%%%%%%%%%%%%%%%%%

\subsection{Hyperbolicity and wave-set structure}
\label{3.2}

%%%%%%%%%%%%%%%%%%%%%%%%%%%%%%%%%%%%%%%%%%%%%%%%

In this section we analyze the hyperbolicity of system (\ref{eq:ffe-eq's-full}). For that, it is enough to study the behavior of high frequency linearized perturbations in off an arbitrary background, \cite{Friedrichs54}. Due to the finite propagation speed of such perturbations, it turns out that we only need to concentrate in a very small neighborhood around an arbitrary point of space-time.

For $\varepsilon > 0$, let us consider the one-parameter family of solutions of (\ref{eq:ffe-eq's-full}) given by:
\begin{equation}\label{high_freq_sol}
\psi_i(\varepsilon) = \phi_i + \varepsilon \; \varphi_i\; e^{\;f/\sqrt{\varepsilon}}\;,
\end{equation}
where $\phi_i$ is any background solution of (\ref{eq:ffe-eq's-full}), and $f$ a smooth complex scalar field. As $\varepsilon$ approaches to zero, more oscillations there will be in a given small neighborhood, and they will become in size closer and closer to the background solution $\phi_i$. We shall refer to this limit as the \textit{high-frequency limit}. This is analog to the limit as the wave number in the space-like plane of an observer $t^a$ tends to infinity.

Replacing (\ref{high_freq_sol}) in (\ref{eq:ffe-eq's-full}) and taking carefully\footnote{Recall that if $p:\mathcal{O}\to\mathbb{C}$ is a complex-valued continuous function defined over a neighborhood $\mathcal{O}\subset\mathcal{M}$, then
\[
\lim_{\varepsilon\to 0}{p(x)e^{\frac{q(x)}{\varepsilon}}}
\]
exists on $\mathcal{O}$ \textit{if and only if} $p\equiv 0$.} the limit $\varepsilon\to 0$, we get the following algebraic equation for $\varphi_i$:
\begin{equation}\label{eq:algebr_EFFsystem}
\varepsilon^{ij} \left(\ell_{ak} \ell^b{}_j - G_{kj}\delta^b{}_a\right) k^a k_b \varphi_i = 0,
\end{equation}
where $\ell^a{}_i$ and $G_{kj}$ are evaluated at the background solution, and $k_a:=\nabla_a f$. 

Equation (\ref{eq:algebr_EFFsystem}) is called the \textit{principal part} of (\ref{eq:ffe-eq's-full}), and contains only the terms of the corresponding linearized equation of (\ref{eq:ffe-eq's-full}) that are of higher orders in frequency. It describes completely the characteristic structure (or wave-set) of the system with respect to a generic wave front propagation plane, $k_a$. 

Let us study the characteristic equations of system (\ref{eq:ffe-eq's-full}). Recall that \textit{real roots} of these equations determine the causal cone structure of the theory, i.e., which are the propagation planes through which wavelike solutions can propagate.

By defining the scalars $\kappa_i := \ell_i \cdot k$, with $i = 1, 2$, equation (\ref{eq:algebr_EFFsystem}) is equivalent to the problem
\begin{equation}\label{eqn:kernel}
\mathcal{A}^i{}_j\; \varphi_i = 0,
\end{equation}
where the operator $\mathcal{A}$ is given by
\begin{equation}
\mathcal{A}^i{}_j := \varepsilon^{i \ell} \left(\kappa_j \kappa_{\ell} - k^2 G_{j \ell}\right),
\end{equation}
and $k^2 = k^a k_a$. Since we are looking for non trivial solutions of (\ref{eqn:kernel}), the \textit{dispersion relation} becomes
\begin{equation}\label{eq:disp_eq}
\det\left(\mathcal{A}\right) = k^2\left[\frac{Fk^2}{2} - G^{ij}\kappa_i \kappa_j\right] = 0.
\end{equation} 
Thus, the possible planes are given by  $k^2 = 0$ or $\frac{F}{2} k^2 - G^{ij} \kappa_i \kappa_j = 0$. Notice that if $\kappa_1 = \kappa_2=0$ (this is the case when $k^a $ is \textit{tangent} to the field sheet) both conditions coincide, and the operator $\mathcal{A}$ is trivial. Thus, $\varphi_i$ is arbitrary and there are two linearly independent solutions for each null direction. 

Otherwise, let us analyze both cases separately:

\begin{itemize}
%%%%%%%%%%%%%%%%%%%%%%%%

\item \textbf{Case I} (\textit{Cone}): $k^2=0$.

If $\kappa_1$ or $\kappa_2$ are not null, then $\varphi_i = \kappa_i$ is a solution of (\ref{eqn:kernel}). Thus, we obtain a 1-dimension space of solutions for each null direction, given by $\{\alpha\kappa_i\;:\;\alpha\in\mathbb{R}\}$.

%%%%%%%%%%%%%%%%%%%%%%%%

\item \textbf{Case II} (\textit{Wedge}): $\frac{F}{2} k^2 - G^{ij} \kappa_i \kappa_j = 0$.

Recalling that $\frac{F}{2} = \det (G)$, $G^{ij}=\varepsilon^{ik} \varepsilon^{j\ell} G_{k \ell}$, and setting $\ell_i := |\ell_i|\; n_i$, with $n_1 \cdot n_1 = n_2 \cdot n_2 = 1$, we get
\begin{eqnarray}
0 & = & \det (G) k^2 - G^{ij} \kappa_i \kappa_j  \nonumber \\
& = & |\ell_1|^2|\ell_2|^2 \left[\left( 1 - (n_1 \cdot n_2)^2\right) k^2 \right. - \left((k\cdot n_1)^2 \right. \nonumber \\
&& + (k\cdot n_2)^2 - \left. \left. 2 (n_1 \cdot n_2)(k\cdot n_1)(k\cdot n_2)\right)\right] \nonumber \\
& = & |\ell_1|^2|\ell_2|^2 \left[1 - (n_1 \cdot n_2)^2\right] \left(k^2 - k^2_{\perp}\right) \nonumber \\
&=& |\ell_1|^2|\ell_2|^2 \left[1 - (n_1 \cdot n_2)^2\right] k^2_{||}\;,
\end{eqnarray}
where $k_{\perp}$ is the norm of the component of $k^a$ in the space spanned by $\ell^a{}_1$ and $\ell^a{}_2$ (i.e., perpendicular to the magnetic sheet). 
Thus, since $(n_1 \cdot n_2)^2 < 1$, the expression vanishes \textit{if and only if} $k = k_{\perp}$.  
Choosing an orthonormal basis $\{e_i^a\}$ in which $\{e_0^a,\; e_3^a\}$ are over the flux surfaces, we get
\begin{equation}\label{eq:norm_paralell}
k^2 = - k_0^2 + k_3^2 + k_{\perp}^2, 
\end{equation}
and we see that the condition for which we have roots is that the wave vector part that is \textit{perpendicular} to $\ell^a{}_i$ must be 
null. Thus, if $k_3 \neq 0$, we have only two possibilities: $k_0 = k_3$ and $k_0 = - k_3$.

On the other hand notice that, by virtue of the invertibility of $G^{ij}$, the problem (\ref{eqn:kernel}) is further equivalent to
\begin{equation}\label{eq:kernel3}
\mathcal{B}^i{}_j\;\varphi^j = 0,
\end{equation}
where the operator $\mathcal{B}$ is given by
\begin{equation}
\mathcal{B}^i{}_j := \tilde{G}^{i\ell} \left(\kappa_j \kappa_{\ell} - k^2 G_{j \ell}\right),
\end{equation}
and $\tilde{G}^{ij}$ given by (\ref{eq:tildeG=G}). Moreover, we get straightforwardly that
\begin{equation}
\mathcal{B}^i{}_j = -k^2\; h^i{}_j,
\end{equation}
where
\begin{equation}\label{eq:proj}
h^i{}_{j} := \delta^{i}{}_j - \frac{\tilde{G}^{i \ell} \kappa_{\ell} \kappa_j}{k^2}
\end{equation}
is a \textit{projector} into the space \textit{perpendicular} to $\kappa_i$ with respect to $G_{ij}$, seeing $G_{ij}$ as an ``internal metric'' with inverse $\tilde{G}^{ij}$ given by (\ref{eq:tildeG=G}). Indeed, $h^i{}_j h^j{}_k = h^i{}_k$, and defining $\bar{\kappa}^i := \tilde{G}^{ij} \kappa_j$ we get 
\begin{equation}
h^i{}_{j}\;\bar{\kappa}^j  = \frac{2\bar{\kappa}^i}{Fk^2} \left(\frac{F}{2}k^2 - G^{\ell m} \kappa_{\ell} \kappa_{m}\right) = 0.
\end{equation}

Thus, $\varphi^i = \bar{\kappa}^i$ is a solution of (\ref{eq:kernel3}), implying that $\phi_i = \varepsilon_{ij}\bar{\kappa}^j$ is a solution of (\ref{eqn:kernel}). We get, again, a 1-dimensional space of solutions for each of the directions obtained above, given by $\{\gamma\varepsilon_{ij}\bar{\kappa}^j\;:\;\gamma\in\mathbb{R}\}$.
\end{itemize}

%%%%%%%%%%%%%%%%%%%%%%%%%%%%%%%%%%%%%%%%%%%%%%%%%%%%%%%%%%%%%%%%

\subsection{Equivalent first order reduction of the algebraic equations}

%%%%%%%%%%%%%%%%%%%%%%%%%%%%%%%%%%%%%%%%%%%%%%%%%%%%%%%%%%%%%%%%

In this section we perform a $3+1$ decomposition of the principal part (eq. \ref{eq:algebr_EFFsystem}) of system (\ref{eq:ffe-eq's-full}) following the guidelines of \cite{Thorne82}. {Then, we reduce it into an equivalent first order system in a very particular way that avoids the appearance of spurious constraints.} Using the information obtained from the wave-set in the previous section, we analyze the kernel of the equivalent reduced system, and see that there is not a complete eigenvector set at each direction.

Let us consider a space-like hypersurface $\Sigma_o$, take an arbitrary point $p\in\Sigma_o$ and choose a gauge transformation like (\ref{eq:gauge_transf}) such that $\ell^a{}_i$ are \textit{perpendicular} to each other on $p$. Let $\mathcal{O}_p$ be an open neighborhood of $p$ within an open set $\mathcal{O}\subset\mathcal{M}$ which is foliated by spacelike hypersurfaces that are the level surfaces of a smooth time function $t:\mathcal{O}\to\mathbb{R}$. Let $t^a := (\partial / \partial t)^a$  be the normal vector field of each hypersurface on the foliation, and choose the coordinate $t$ such that $t^a \nabla_a t = 1$. Over each hypersurface, define an orthonormal frame $\{e^a{}_i\}_{i=0}^3$, with $e^a_0 = (\partial / \partial t)^a$ and for $i=1,2$, $e^a{}_i$ are along the $\ell^a{}_i$ direction. 

Using the customary notation
\[
\ell^a{}_i = (0,\vec{\ell}_i), \quad k^a = (k_0, \vec{k}), \;\;\ell^a{}_j\; k_a = \vec{\ell}_j \cdot \vec{k},
\] 
equation (\ref{eq:algebr_EFFsystem}) reads
\begin{equation} \label{eq:3+1_gral_syst}
- \varepsilon^{ij} G_{kj} k^2_0 \varphi_i + \varepsilon^{ij} \left[|\vec{k}|^2 G_{kj} - (\vec{\ell}_k\cdot\vec{k}) (\vec{\ell}_j\cdot\vec{k})\right] \varphi_i=0.
\end{equation}
By defining the variables
\begin{equation}
u_i = k_0 \varphi_i\;; \quad v_i = |\vec{k}|\varphi_i\;,
\end{equation}
we obtain the following system:
\begin{equation} \label{eq:total_matrix}
\left( \begin{array}{cccc}
k_0 & 0 & - \frac{k_1^2 + k_3^2}{|\vec{k}|} & - \frac{(\vec{\ell}_1\cdot\vec{k})(\vec{\ell}_2\cdot\vec{k})}{|\vec{k}| G_{22}} \\
0 & k_0 & - \frac{(\vec{\ell}_1\cdot\vec{k})(\vec{\ell}_2\cdot\vec{k})}{|\vec{k}| G_{11}} & - \frac{k_2^2 + k_3^2}{|\vec{k}|} \\
|\vec{k}| & 0 & -k_0 & 0 \\
0 & |\vec{k}| & 0 & -k_0 \end{array} 
\right)
\left( \begin{array}{cccc}
u_1 \\
u_2 \\
v_1 \\
v_2 
\end{array} 
\right) = 0.
\end{equation}
System (\ref{eq:total_matrix}) is \textit{completely equivalent} to (\ref{eq:3+1_gral_syst}) in the following sense: for any $\vec{k} \neq 0$, there exists a biunivocal relation between solutions of both systems (\ref{eq:total_matrix}) and (\ref{eq:3+1_gral_syst}); that is, every solution of the original second order system, (\ref{eq:3+1_gral_syst}), is a solution of the above system and vice-versa. In particular, this method of obtaining first order systems out of second order in Fourier space does not include any constraint nor spurious solutions on the initial data, and that is why both systems are equivalent as well. Thus, if we show that the above first order system is ill-posed, the original second order system will also be ill-posed.

It is straightforward to check that the matrix in (\ref{eq:total_matrix}) is generally non diagonalizable when $k_0 = k_3 = 0$. We shall later exhibit this feature by choosing particular configurations for $k^a$ and $\ell^a{}_i$, from which we shall find solutions that grow linearly in frequencies.

The solutions of the above system, both in the Cone-case and in the Wedge-case exhibited in the previous section, are given by
\begin{equation}\nonumber
\mbox{\textit{Cone:}} \;\; u_i  =  k_0\; \kappa_i\;\;,\;\; v_i = |\vec{k}|\;\kappa_i\;;
\end{equation} 
\begin{equation}\label{eq:sol_kernel_nuevas}
\mbox{\textit{Wedge:}} \;\; u_i  =  - k_0\; \varepsilon_{ij}\tilde{G}^{j \ell}\kappa_{\ell}\;,\; v_i = - |\vec{k}|\;\varepsilon_{ij}\tilde{G}^{j \ell}\kappa_{\ell}\;.
\end{equation} 

Thus, if $k_0 \neq 0$, we have in total \textit{four} linearly independent solutions of (\ref{eq:total_matrix}) when considering both Cone-case (that is, $k_0 = \pm |\vec{k}|$) and Wedge-case ($k_0 = \pm k_3$).

On the contrary, when $k_0 = k_3 = 0$ (that is, at the edge of the wedge), we only get one solution in this case, getting in total \textit{three} linearly independent solutions.

In summary, we have found a complete set of solutions of (\ref{eq:total_matrix}) at all points except at the points lying in the straight line $k_0 = k_3 = 0$. This is the case when the wave-vector is \textit{perpendicular} to the plane $\{e_0,e_3\}$, so $k^a = \mbox{span}\{\ell_1,\;\ell_2\}$; and we get
\begin{equation} \label{eq:k_par}
k^a=\bar{\kappa}^i \ell^{a}_i.
\end{equation}

It is interesting to note that perturbations of $F^{ab}$ constructed from those perturbations of $\phi_i$ for which $k_0 = k_3 = 0$, vanish. Indeed, recall that a general perturbation of Maxwell's tensor written like in (\ref{eq:max_pot_sympl}) is given by
\begin{equation}
\delta F^{ab} = 2\varepsilon^{ij}\ell^{[a}{}_i X^{b]}{}_j,
\end{equation}
where $X^a{}_i = \delta\ell^a{}_i$. In the high frequency limit, we get $X^a{}_i = \varphi_i k^a$, where $\varphi_i = \delta\phi_i$. Using it in the above formula for $\delta F^{ab}$, we get
\begin{equation}
\delta F^{ab} = 2\varepsilon^{ij}\varphi_j\ell^{[a}{}_i k^{b]}.
\end{equation}
Setting $\varphi^i = \bar{\kappa}^i$ and $k^a$ given in (\ref{eq:k_par}), we see that $\delta F^{ab} = 0$. Thus, whenever $k_0 = k_3 = 0$, perturbations of $F^{ab}$ constructed from non vanishing perturbations of $\ell^a{}_i$ vanish. These are \textit{spurious modes} because they do not appear when considering just Maxwell perturbations, and provide gauge solutions that might make the system to be ill posed, as we shall see in the next section.
$\vspace{0.7cm}$
%%%%%%%%%%%%%%%%%%%%%%%%%%%%%%%%%%%%%%%%%%%%%%%%

\section{Ill posedness}
\label{IV}

\subsection{Kreiss's algebraic criterion failure}

%%%%%%%%%%%%%%%%%%%%%%%%%%%%%%%%%%%%%%%%%%%%%%%%

Strang's theorem asserts that if a quasi-linear system like (\ref{eq:strang-syst}) is well-posed, then the system that results by evaluating the variable-coefficients in any point is also well-posed. Thus, to show that the present system is not well-posed, it suffices to check that at least one of its constant coefficients renditions is ill-posed. Recall that Kreiss's algebraic criterion provided in theorem \ref{kreiss} (see section \ref{s:kreiss_theorem}) is valid only for constant-coefficient first order systems. Thus, in order to apply it, let us assume that the linearization procedure done in (\ref{high_freq_sol}) is around background solutions with \textit{constant} $\mathring{\ell}^a{}_i$. 

For simplicity, let us choose a particular configuration of system (\ref{eq:total_matrix}), such that $k_1=k_2=\sqrt{2} \kappa$, $0 < \kappa \in\mathbb{R}$, and $k_0 = k_3 = 0$. Now, for $s\in\mathbb{R}$ and following Kreiss's theorem (see Theorem \ref{kreiss}), let us construct the matrix $\mathcal{D}:=\mathbb{A}-sI$ given by

\begin{equation} \label{eq:matrix_example1}
\mathcal{D}= 
\left( \begin{array}{cccc}
-s & 0 & -\kappa & -\alpha \kappa \\
0 & -s & -\kappa/\alpha & -\kappa \\
2\kappa & 0 & -s & 0 \\
0 & 2\kappa & 0 & -s \end{array} 
\right),
\quad
\alpha:=\frac{|\ell_1|}{|\ell_2|} > 0\;,
\end{equation}
where $\mathbb{A}$ is the matrix of (\ref{eq:total_matrix}) evaluated in this particular configuration. The inverse of $\mathcal{D}$ is computed to be
\begin{widetext}
\begin{equation} \label{eq:inverse_ex1}
\mathcal{D}^{-1}=\frac{1}{s^2\left(s^2 + 4\kappa^2\right)} 
\left( \begin{array}{cccc}
-s\left(s^2 + 2\kappa^2\right) & -2\alpha \kappa^2 s & \kappa s^2 & -\alpha \kappa s^2 \\
\frac{2\kappa^2s}{\alpha} & s\left(s^2 + 2\kappa^2\right) & \frac{\kappa s^2}{\alpha} & -\kappa s^2 \\
-2\kappa \left(s^2 + 2\kappa^2\right) & -4\alpha \kappa^3 & -s\left(s^2 + 2\kappa^2\right) & -2\alpha \kappa^2 \\
\frac{4\kappa^3}{\alpha} & 2\kappa\left(2\kappa^2-s^2\right) & \frac{2\kappa^2s}{\alpha} & s\left(s^2 + 2\kappa^2\right) \end{array} 
\right).
\end{equation}
\end{widetext}

Note that there are elements of the above matrix that cannot be bounded like the resolvent condition of Kreiss's Theorem. It is evident\footnote{Indeed,
\[
|(\mathbb{A}-sI)^{-1}_{32}| = \frac{4\alpha \kappa^3}{s^2(s^2 + 4\kappa^2)} > \frac{\beta}{s}
\]
if and only if
\[
p(s):= -\beta s^3 - 4\beta \kappa^2 s + 4\alpha \kappa^3 > 0.
\]
Since $p(0)=4\alpha \kappa^3 > 0$, continuity of $p$ guarantees an open interval $I\subset\mathbb{R}$ of positive values for $s$ for which $p(s)>0$.} that there exists an open interval $I\subset\mathbb{R}$ of positive values of $s$ such that, for all $\beta > 0$,
\begin{equation} \label{eq:div}
|(\mathbb{A}-sI)^{-1}_{32}| > \frac{\beta}{s}.
\end{equation}
Thus, the resolvent condition does not hold for all $s\in\mathbb{C}$, and system (\ref{eq:total_matrix}) is not well posed. 

We recall here that since Kreiss's criterion is applicable to (constant) matrices, it can be used only for \textit{constant-coefficients} linear first order systems. Nevertheless, the generalization to the quasi-linear first order system is direct, using the result provided by Strang in \cite{Strang66}. Strang asserts that if a quasi-linear first order system is well posed, then the linear system obtained by freezing the coefficients at any arbitrary point is also well posed. Thus, the counter-reciprocal statement leads us to conclude the opposite: since the system (\ref{eq:total_matrix}) with frozen coefficients is not well posed, then the general quasi-linear system shares the same property.

%%%%%%%%%%%%%%%%%%%%%%%%%%%%%%%%%%%%%%%%%%%%%%%%
%\vspace{-1.1cm}
\subsection{Constructing diverging initial data}

%%%%%%%%%%%%%%%%%%%%%%%%%%%%%%%%%%%%%%%%%%%%%%%%

As a consequence of what we have shown in the previous section, we should find solutions of the system that grow in frequency and time. The divergence with $s$ in (\ref{eq:div}) was of second order, so we expect to have a Jordan block of order one (i.e., only one missing eigenvector) and so a mode growing linearly both in frequency and time. It is of our interest to display their behavior because they may eventually appear in numerical simulations. In the next section, we will use this mechanism to generate an explicit bounded sequence (in Sobolev norms) of initial data such that the corresponding evolution sequence diverges.

By the identification $k_0 \leftrightarrow i\partial_t$ in (\ref{eq:total_matrix}), we arrive to the linear first order system given by 
\begin{equation} \label{eq:eq's_1st_or_sys}
\partial_t U = \mathbb{A}\;U, \qquad \mbox{where}
\end{equation}
\begin{equation} \label{eq: 1st_matrix_syst}
\mathbb{A}= 
\left( \begin{array}{cccc}
0 & 0 & \frac{-k_1^2-k_3^2}{|\vec{k}|} & -\frac{(\vec{\ell}_1\cdot\vec{k})(\vec{\ell}_2\cdot\vec{k})}{|\vec{k}| G_{22}} \\
0 & 0 & -\frac{(\vec{\ell}_1\cdot\vec{k})(\vec{\ell}_2\cdot\vec{k})}{|\vec{k}| G_{11}} & \frac{-k_2^2-k_3^2}{|\vec{k}|} \\
|\vec{k}| & 0 & 0 & 0 \\
0 & |\vec{k}| & 0 & 0 \end{array} 
\right)\;,
\end{equation}
and we have redefined the variables such that 
\begin{equation}
U=\left( \begin{array}{cc}
 \partial_t \hat{\varphi}_i \\
 |\vec{k}|\;\hat{\varphi}_i\end{array} \right), \quad
 i = 1,2;
\end{equation}
where we have also identified $\varphi_i\leftrightarrow\hat{\varphi}_i$. System (\ref{eq:eq's_1st_or_sys})
\footnote{A ``differential'' way to get (\ref{eq:eq's_1st_or_sys}) is by taking a linearization of the full system (\ref{eq:ffe-eq's-full}) around a solution $\phi^o{}_i$ with constant $\mathring{\ell}^a{}_i$, and Fourier transform in space.}
will be used to find explicit solutions that diverge in frequency and time, and to provide a sequence of initial data for which continuity of the evolution with respect to it does not hold.

The matrix $\mathbb{A}$ in (\ref{eq: 1st_matrix_syst}) has four imaginary eigenvalues given by
\[
\lambda^{(1)}_{\pm}(\vec{k}) = \pm\frac{|k_3|}{\sqrt{2}}\;i, \quad \lambda^{(2)}_{\pm}(\vec{k}) = \pm i \sqrt{k_1^2 + k_2^2 + \frac{k_3^2}{2}},
\]
and in particular Re$\left(\lambda^{(j)}_{\pm}\right)=0$. Thus, we can expect a priori the system to be well-posed\footnote{See \cite{Kreiss89}, Lemma 2.3.1.}. Nevertheless, if $k_3=0$, then $\lambda^{(1)}=0$ is a root with multiplicity 2, and if the system were well-posed, the corresponding Jordan block of the Jordan matrix $J=P^{-1}\mathbb{A}P$ should be diagonal, that is, the corresponding eigenspace should have dimension 2. 

Nevertheless, for $\vec{k}\neq 0$, there is a unique form to have an eigenvalue with algebraic multiplicity greater than 1, that is, taking $k_3=0$. In that case, the eigenvalues are
\begin{equation} \label{eq:eigenvalues}
\lambda_{\pm} = \pm\; i\; |\vec{k}|, \qquad \lambda_0 = 0,
\end{equation}
and a Jordan decomposition of $\mathbb{A}$ is
\begin{equation} \label{eq:J_and_P_matrix}
J = 
\left( \begin{array}{cccc}
i|\vec{k}| & 0 & 0 & 0 \\
0 & -i|\vec{k}| & 0 & 0 \\
0 & 0 & 0 & |\vec{k}| \\
0 & 0 & 0 & 0 \end{array} 
\right),
\end{equation}
which clearly has a missing eigenvector. Exponenciating $\mathbb{A}$ using the above Jordan decomposition, we get the general solution of (\ref{eq:eq's_1st_or_sys}):
\begin{widetext}
\begin{equation} \label{eq:general_solution_explicit}
U(t) =
\left( \begin{array}{cccc}
i e^{i|\vec{k}|t} & -i e^{-i|\vec{k}|t} & 0 & -\frac{|\ell_1|k_2}{|\ell_2| k_1} \\
i e^{i|\vec{k}|t} & -ie^{-i|\vec{k}|t} & 0 & 1 \\
e^{i|\vec{k}|t} & e^{-i|\vec{k}|t} & -\frac{|\ell_1|k_2}{|\ell_2| k_1} & -\frac{|\ell_1|k_2}{|\ell_2|k_1} (1+|\vec{k}| t) \\
e^{i|\vec{k}|t} & e^{-i|\vec{k}|t} & 1 & 1+|\vec{k}| t \end{array} 
\right)
\left( \begin{array}{cccc}
 V^{0}_1 \\
 V^{0}_2 \\
 V^{0}_3 \\
 V^{0}_4 \end{array} \right).
\end{equation}
\end{widetext}

From here we see that choosing any set $(V_1^{0},\cdots,V_4^{0})\in\mathbb{R}^4$ with $V_4^{0}\neq 0$, we generate initial data that give rise to solutions that \textit{grow linearly} both in frequency and time.

%%%%%%%%%%%%%%%%%%%%%%%%%%%%%%%%%%%%%%%%%%%%%%%%

\subsection{On the lack of continuity along evolution}

%%%%%%%%%%%%%%%%%%%%%%%%%%%%%%%%%%%%%%%%%%%%%%%%

Due to the fact that there are solutions that grow linearly in $|\vec{k}|$, it is possible to see that the evolution (\ref{eq:general_solution_explicit}) is in general non continuous with respect to the initial data. To see this, it suffices to fix an instant of time $t=T>0$, (which could be taken arbitrarily small) and check that there does not exist a constant $C>0$ such that
\begin{equation}
\norm{u(T,x)} \leq C \norm{f(x)},
\end{equation}
for all initial data $f(x)$ given at $t=0$, where $u(T,x)$ is the corresponding evolution of that data until $t=T$.

To see that such $C$ does not exist we shall construct a bounded sequence of initial data for (\ref{eq:eq's_1st_or_sys}), such that the corresponding sequence of solutions is unbounded in norm at time $T$, (from now on, we will refer that sequence as the \textit{evolution sequence}). 
We shall build this sequence using the same configuration as in (\ref{eq:matrix_example1}), in which $k_1 = k_2$, $k_3 = 0$, and $\alpha:=|\ell_1|/|\ell_2|$.

%\norm{U(t,k)}_{2} := \sqrt{\abs{\partial_t \phi_1}^2 + \abs{\partial_t \phi_2}^2 + |\vec{k}|^2\abs{\phi_1}^2 + |\vec{k}|^2\abs{\phi_1}^2}}\;.

To build the solution we shall use the finite propagation speed of perturbations (which is the speed of light) so that we can consider \textit{locally plane wave} solutions.
Consider a flat background spacetime which is foliated by constant time planes. 
Let $\Sigma_o$ be the slice $\{t=0\}$, and for $R,T>0$, let $\mathcal{B}(R,T)\subset\Sigma_o$ be the ball of radius $R+T$. 
Let us consider  a smooth background solution $\phi_i$ such that on the domain of dependence of $\mathcal{B}(R,T)$ the gradients $\ell^a{}_i$ are constant and perpendicular to each other. 
Suppose that outside the ball, background solutions decay smoothly to zero so that the corresponding norms are uniformly bounded.

We shall be looking for initial data
\begin{equation}
\Phi_o = (\varphi_1^o, \varphi_2^o), \quad \partial_t\Phi|_o = ((\partial_t\varphi_1)^o, (\partial_t\varphi_2)^o)
\end{equation}
of (\ref{eq:eq's_1st_or_sys}) such that $\Phi_o\in H^1(\Sigma_o,\mathbb{R}^2)$ and $\partial_t\Phi|_o \in L^2(\Sigma_o,\mathbb{R}^2)$. 
As usual, define the norm of the solution $\Phi(t) = (\varphi_1, \varphi_2)$ at time $t$ as

\begin{equation}\label{eq:norm-H1}
\norm{\Phi(t)} := \left(\norm{\partial_t\Phi}_{L^2(\Sigma_t,\;\mathbb{R}^2)}^2 + \norm{\Phi}_{H^1(\Sigma_t,\;\mathbb{R}^2)}^2\right)^{1/2},
\end{equation}
where $\norm{\Phi}_{L^2(\Sigma,\mathbb{R}^2)} := \norm{\abs{\Phi}}_{L^2(\Sigma)}$ and
\begin{equation}
\norm{\Phi}_{H^1(\Sigma,\mathbb{R}^2)}^2:= \int_{\Sigma}{\abs{\varphi_1}^2 + \abs{\varphi_2}^2 + \abs{\nabla\varphi_1}^2 + \abs{\nabla\varphi_2}^2}.
\end{equation}

Let us consider the following initial data sequence for the perturbations, given on $\Sigma_o$:
\begin{equation}
\varphi^n_1|_o = \varphi^n_2|_o = 0;
\end{equation}
\begin{equation}
{(\partial_t\varphi^n_1)|_o(x)=\left\{
{\begin{array}{rcl}
e^{i k_n\cdot x}/\sqrt{n},\;\mbox{if}&x&\in \mathcal{B}(R,T)\\
g_n(x),\; \mbox{if}&x&\in\Sigma_o \setminus \mathcal{B}(R,T)
\end{array}}
\right.}
\end{equation}
\begin{equation}
(\partial_t\varphi^n_2)|_o = -\frac{(\partial_t\varphi^n_1)|_o}{\alpha};
\end{equation}
where $k_n = n(1,1,0)$ and $g_n$ is a bounded sequence in $L^2(\Sigma_o\setminus \mathcal{B}(R,T))$ such that $\partial_t\phi^n_1|_o$ is smooth over $\Sigma_o$ and the norm of each element of the initial data sequence is bounded.

Denoting $\Phi^n := (\varphi^n_1,\; \varphi^n_2)$, the norm of the above data is 
\begin{align}
\norm{\Phi^n_o}^2 & = \norm{\partial_t\Phi^n_o}^2_{L^2(\Sigma_o)} \nonumber \\
& =  \norm{\partial_t\Phi^n_o}^2_{L^2(\mathcal{B}(R,T))} + \norm{\partial_t\Phi^n_o}^2_{L^2(\Sigma_o\setminus \mathcal{B}(R,T))} \nonumber \\
& =  \left(1+\frac{1}{\alpha^2}\right)\left[\frac{\abs{\mathcal{B}(R,T)}}{n} + \norm{g_n}^2_{L^2(\Sigma_o\setminus\mathcal{B}(R,T))}\right]
\end{align}
where $\abs{\mathcal{B}(R,T)}$ is the volume of the ball in $\Sigma_o$. Thus, this sequence is uniformly bounded since $g_n$ is so.

We shall see now that the restriction of the corresponding evolution sequence on the ball $\mathcal{B}(R)\subset\Sigma_T$ of radius $R$ at $\Sigma_T = \{t=T\}$ grows without bound. Thus, if the restriction grows with $n$ without bound, the norm over the full time slice $\Sigma_T$ will also will also grow without bound. 
Finite propagation speed ensures us that the solution sequence at $\mathcal{B}(R)$ will be just the evolution of the original plane wave sequence defined on $\mathcal{B}(R,T)$. 
Indeed, this solution will be unique in the whole domain of dependence of $\mathcal{B}(R,T)$ as a consequence of Holmgren's uniqueness theorem\footnote{See \cite{Smoller83}, Theorem 5.1., and \cite{Reula04}, Prop. 5. for references.}, since the corresponding initial data is smooth and is given over $\mathcal{B}(R,T)\subset\Sigma_o$ which is a non-characteristic surface. 
Thus, in order to analyze the behavior of the evolution on $\mathcal{B}(R)$, it suffices to evolve the restriction of the initial data on $\mathcal{B}(R,T)$ as if it were a plane wave over the whole space. This corresponds to evolve system (\ref{eq:eq's_1st_or_sys}) by taking as initial data
\begin{equation}
\hat{\varphi}^n_i|_o = 0, \; \partial_t\hat{\varphi}^n_1|_o = \frac{1}{\sqrt{n}}, \; \partial_t\hat{\varphi}^n_2|_o = -\frac{\partial_t\hat{\varphi}^n_1|_o}{\alpha}.
\end{equation}
The evolution at time $t=T$ is given by
\begin{equation}
\hat{\varphi}^n_1(T,\vec{k}) = \frac{T}{\sqrt{n}}, \quad
\hat{\varphi}^n_1(T,\vec{k}) = -\frac{T}{\alpha\sqrt{n}},
\end{equation}
which corresponds to the \textit{unique} solution
\begin{equation}\label{eq:sol}
\varphi^n_1(T,x) = \frac{T}{\sqrt{n}}e^{i k_n\cdot x}; \quad \varphi^n_2(T,x) = -\frac{T}{\alpha\sqrt{n}}e^{i k_n\cdot x},
\end{equation}
inside $\mathcal{B}(R)\subset\Sigma_T$. 

For the norm of the full solution at time $t=T$, we get
\begin{align}
\norm{\Phi^n(T)}^2 & \geq \norm{\Phi^n(T)|_{\mathcal{B}(R)}}^2 \nonumber \\
& = \norm{\Phi^n(T)}_{H^1(\mathcal{B}(R))}^2 + \norm{\partial_t\Phi^n(T)}_{L^2(\mathcal{B}(R))}^2 \nonumber \\
& = \norm{\Phi^n(T)}_{L^2(\mathcal{B}(R))}^2 + \norm{|\nabla\Phi^n(T)|}_{\mathcal{B}(R))}^2 \nonumber \\ 
& \hspace{3.3cm} +\norm{\partial_t\Phi^n(T)}_{L^2(\mathcal{B}(R))}^2 \nonumber \\
& =  2T^2\left(1 + \frac{1}{\alpha^2}\right)\abs{\mathcal{B}(R)}n + \mathcal{O}(1/n),
\end{align}
where we have denoted
\[
\norm{|\nabla\Phi|}_{\mathcal{B}(R))}^2 := \sum_{j=1}^2{\norm{\abs{\nabla\varphi_i}}_{L^2(\mathcal{B}(R))}^2}.
\]
Thus, there can not exist a bound of the solution in terms of a bound of the initial data, for any finite time, $t$.

A similar proof is also valid for any Sobolev norm,  thus controlling an arbitrary finite number of derivatives, and moreover, as shown by Strang, it can be extended for perturbations around arbitrary smooth solutions. Essentially, as we are considering perturbations of higher frequencies, we can zoom in to smaller neighborhoods. Assuming the background solution to be smooth, it only matters their values at the zooming points.

%%%%%%%%%%%%%%%%%%%%%%%%%%%%%%%%%%%%%%%%%%%%%%%%

\subsection{Ill posedness in the Leray-Ohya sense}

%%%%%%%%%%%%%%%%%%%%%%%%%%%%%%%%%%%%%%%%%%%%%%%%

Leray-Ohya hyperbolicity \cite{Leray53, Leray-Ohya67} seems to be a weaker condition than strong hyperbolicity, for it uses topologies which do not arise from norms, but rather from more general topological spaces, as Gevrey spaces, where \textit{semi-norms} weighting derivatives of functions to all orders are used (see \cite{Choquet82} for detailed discussions). Thus, one might entertain the idea that Force-free in Euler potentials could be hyperbolic in that sense. We show here that this is not the case. We refer the reader to Appendix \ref{app1} for notations and definitions.

We begin by considering the system (\ref{eq:ffe-eq's-full}) which is a set of two partial differential equations for the potentials $\phi_1$ and $\phi_2$. This system can be put in the Leray form (\ref{eq:leray-general}). Indeed, setting $N=2$, $u^1 = \phi_1$ and $u^2 = \phi_2$, we get
\begin{equation} \label{eq:1}
H^1{}_1 \equiv (\nabla_a\phi_1)(\nabla^a\phi_2)\nabla^b\nabla_b - (\nabla_a\phi_1)(\nabla^b\phi_2)\nabla^a\nabla_b\;;
\end{equation}
\begin{equation}
H^1{}_2 \equiv (\nabla_a\phi_1)(\nabla^b\phi_1)\nabla^a\nabla_b - (\nabla_a\phi_1)(\nabla^a\phi_1)\nabla^b\nabla_b\;;
\end{equation}
\begin{equation}
b^1 = 0\;;
\end{equation}
and similarly for the second equation. By this way, (\ref{eq:ffe-eq's-full}) now reads
\begin{equation} \label{eq:EC1}
H^1{}_1u^1 + H^1{}_2u^2 = 0\;;
\end{equation}
\begin{equation} \label{eq:EC2}
H^2{}_1u^1 + H^2{}_2u^2 = 0\;;
\end{equation}
and the associated Leray indices are
\begin{equation}
m(\phi_1) = 2;\; m(\phi_2) = 2;\; n(\ref{eq:EC1}) = 0;\; n(\ref{eq:EC2}) = 0.
\end{equation}
By identifying $\nabla_a\leftrightarrow k_a$ in (\ref{eq:EC1})-(\ref{eq:EC2}) and computing the characteristic determinant of the principal part, we arrive to the same dispersion relation obtained in section \ref{3.2}, equation (\ref{eq:disp_eq}), when analyzing the characteristic structure of Force-free systems. While $k^2$ is a hyperbolic polynomial of second degree, it is easy to see that $\frac{Fk^2}{2} - G^{ij}k^a k^b \nabla_a\phi_i \nabla_a\phi_j$ is not hyperbolic.

%%%%%%%%%%%%%%%%%%%%%%%%%%%%%%%%%%%%%%%%%%%%%%%%

\section{Concluding remarks}
\label{V}

%%%%%%%%%%%%%%%%%%%%%%%%%%%%%%%%%%%%%%%%%%%%%%%%

In this article we considered the equations that describe Force-free Electrodynamics using Euler Potentials. Studying the hyperbolicity of such formulation, we found that in this variables the theory is not strongly hyperbolic, and thus the system does not constitute a well-posed initial value problem. This implies that there is no energy (norm) for which the solution is bounded by the same norm in the initial data. To show this, it was sufficient to find an equivalent first order reduction of the equations that violates an algebraically equivalent criterion for strongly hyperbolic systems proposed by Kreiss \cite{Kreiss89}. Using Strang's theorem for constant-coefficient systems we could see that the system is not well posed in general.

We performed a characteristic decomposition of the FFE system in Euler potentials with respect to a generic wave front propagation direction and we derived the resulting causal structure, finding two possible propagation planes. We could find a complete set of eigenvectors in both cases, except at a two dimensional set of planes formed by the intersection of two null planes. This property does not appear when studying of the hyperbolicity of the system in Maxwell variables, see for instance \cite{Pfeiffer13, Carrasco16}. The reason for the occurrence of this peculiarity is that perturbations leading to divergent solutions in the present formulation are not physical, i.e., the Maxwell tensor  $F_{ab}$ constructed from these growing perturbations vanish identically.

On the other hand, and with the aim of displaying the growing modes, explicit initial configurations were constructed such that the subsequent evolution led into fields that developed a linear growth with frequencies. The study of these solutions was completed by showing that evolution is generally non continuous with respect to the initial data. This is so for any norm built out from the initial data and a finite number of its derivatives. To this end we explicitly constructed a bounded initial data sequence and show that the corresponding sequence of solutions at any given time, however small, is unbounded, thus violating continuity for those norms.

Furthermore, the same system was studied in the context of Leray-Ohya hyperbolicity. This kind of hyperbolicity is weaker than the one studied previously (strong hyperbolicity), because it focuses on the initial value problem from initial data that belong to certain spaces of functions whose topologies do not arise from any norm. An example of these spaces are Gevrey classes; i.e., $C^{\infty}$ functions but with Taylor series not necessarily convergent \cite{Choquet82, Choquet08}. We could see that FFEEP is also ill-posed in the sense of Leray-Ohya.

From the above results we conclude that FFEEP should not be used in numerical simulations or other kinds of approximations. Growing linear perturbations will become arbitrarily stiff as the grid frequency is increased. Furthermore non-linearities can alter that growth making it to become exponential, rendering computations nonsensical. 

The above results might not be conclusive in the following sense. There are very simple examples (see, e.g.,\cite{Kreiss89}) for which by choosing Sobolev norms of different weights for different variables one can show continuity. These are very special cases, for generic lower order perturbations of such a systems render them discontinuous. It might be that the present system falls in that category. For those, the general theory is hopeless and so one should aim for finding very particular energy norms and showing their corresponding non-linear estimates. After that, numerical schemes should be used such that those estimates are preserved at the discrete level.

%%%%%%%%%%%%%%%%%%%%%%%%%%%%%%%%%%%%%%%%%%%%%%%%

\section{Acknowledgments}

%%%%%%%%%%%%%%%%%%%%%%%%%%%%%%%%%%%%%%%%%%%%%%%%

We especially want to thank Fernando Ábalos, Federico Carrasco and Miguel Megevand for ideas and discussions throughout this work. This project was partially supported by grants PIP 2014-2016 GI, No. 11220130100470CO and PICT No. 30720150100144CB of SeCyT, UNC. M.E.R. has a doctoral fellowship of CONICET, Argentina.

\appendix

\section{Hyperbolicity in the Leray-Ohya sense}
\label{app1}
\setcounter{equation}{0}

In this appendix we perform a brief review of Leray systems and the notion of hyperbolicity in the Leray-Ohya sense, after introducing the notion of hyperbolic polynomials and hyperbolic operators. We refer the reader to the books \cite{Choquet82, Choquet08}, in which there is a detailed discussion of the original works \cite{Leray53, Leray-Ohya67}.
\\

\textit{Hyperbolic polynomials.} Over a smooth manifold $M$, let $P:T_p^{*} M \to \mathbb{R}$ be a polynomial of degree $n$, with $p\in M$. Let us now consider the set $\mathcal{C}^{*}(p,P):=\{X \in T_p^{*} M\;|\; P(X) = 0\}$. In many contexts, and depending on the particular polynomial $P$ one is considering, $\mathcal{C}^{*}(p,P)$ can be interpreted as the boundary of a \textit{cone} of covectors on $T_p^{*}M$, sometimes used in general relativity for referring to the set of covectors that make positive definite a certain symmetric structure (constructed from $P$, for instance)\footnote{Actually, it is possible to construct covector cones as \textit{duals} to the well-known mathematical cones in the following way: given a (formal) cone $C\subset V$ on a vector space $V$, we define the set $C^{*}:=\{\omega\in V^{*}\;|\;\omega(v)\geq 0,\;\forall v\in C\}$, that is clearly a convex cone.}. The polynomial $P(X)$ is said to be \textit{hyperbolic} if there exists $Y\in T^* M$ such that every straight line passing by $Y$ which does not intercept the origin $X=0$, intersects the set $\mathcal{C}^{*}(p,P)$ in $n$ distinct points. An operator $\mathcal{L}$ is said to be \textit{hyperbolic} at $p$ if it principal part defines a hyperbolic polynomial.
\\

\textit{Leray-Ohya hyperbolicity.} Consider now a system of $N$ partial differential equations for $N$ unknown scalar fields, $u^A$, $A = 1, \cdots, N$, defined over $M$. We say that such system is a \textit{Leray system} if it is possible to associate to any field $u^A$ an non-negative integer $m_I$, $I=1,\cdots, N$ and to each equation another non-negative integer, namely $n_J$, $J=1,\cdots, N$ such that it reads
\begin{align} \label{eq:leray-general}
H^J{}_I (x, & \partial^{m_K - n_J - 1}u^K, \partial^{m_I - n_J})u^I \nonumber \\
& + b^J (x,\partial^{m_K - n_J - 1}u^K) = 0, \;\; J=1,\cdots, N,
\end{align}
where summation over index $I$ is understood and there is not a sum over integers $m_I$ and $n_J$. The operator $H^J{}_I$, known as the \textit{principal part} of (\ref{eq:leray-general}), is an operator of order at most $m_I - n_J$, and it depends on, at most, $m_K - n_J - 1$ derivatives of each field $u^K$. If $m_I - n_J < 0$, then we set $H^J{}_I = 0$. Similarly, if $m_K - n_J < 0$ for some $K$, then $H^J{}_I$ does not depend on $u^K$. The remaining terms $b^J$ also depend on at most $m_K - n_J - 1$ derivatives of each $u^K$, and do not depend on those $u^K$ such that $m_K - n_J < 0$. It is clear that the operator $H^J{}_I$ may not be linear in the fields nor in their derivatives. The numbers $m_I$ and $n_J$ are called \textit{Leray indices}.

Recall that to a given differential operator $\partial^{\alpha}$, we can associate a monomial $k^{\alpha}$ by the way
\[
\partial^{\alpha}:=\partial_0^{\alpha_0}\cdots\partial_n^{\alpha_n} \; \leftrightarrow \; k^{\alpha}:= k_0^{\alpha_0}\cdots k_n^{\alpha_n},
\]
where $k_j^{\alpha_j}$ are real variables. With the above identification, it makes sense to define the \textit{characteristic determinant} of system (\ref{eq:leray-general}), given by
\begin{equation} \label{eq:char-pol}
\mathcal{D}(x,u,k) := \det\left(H^J{}_I(x,\partial^{m_K - n_J - 1}u^K,k^{m_I - n_J})\right).
\end{equation}
This is an homogeneous polynomial of degree $\sum_I{m_I} - \sum_J{n_J}$. If $\mathcal{D} \not\equiv 0$, then we say the system is \textit{regular} in the \textit{Cauchy-Kovalevskaya} sense. Consider now the Cauchy problem associated with (\ref{eq:leray-general}) with initial data given over a Cauchy surface $\Sigma$. We say that system (\ref{eq:leray-general}) is \textit{hyperbolic in the sense of Leray-Ohya} if it is possible to write the characteristic determinant as a product of $q$ hyperbolic polynomials
\begin{equation}
\mathcal{D}(x,u,k) = P_1(x,u,k)\cdots P_q(x,u,k),
\end{equation}
such that the following condition holds:
\begin{equation}
\max_{i}{\{\mbox{deg}(P_i)\}} \geq \max_{I}{\{m_I\}} - \min_{J}{\{n_J\}},
\end{equation}
where $\mbox{deg}(P_i)$ is the degree of $P_i(x,u,k)$. Systems which are hyperbolic in the Leray-Ohya sense are well posed in certain Gevrey class spaces.

%\section{References}

%\subsection{}

%\bibliographystyle{alpha}
%\bibliography{biblio_fluidos_luis}

\begin{thebibliography}{99}

\bibitem{Gralla14}
Samuel E Gralla and Ted Jacobson. 
Spacetime approach to force-free magnetospheres. 
Monthly Notices of the Royal Astronomical Society, 
445(3):2500–2534, 2014.

\bibitem{UchidaI97}
Toshio Uchida. 
Theory of force-free electromagnetic fields. I. General Theory. 
Phys. Rev. E, 56:2181–2197, 1997.

\bibitem{UchidaII97}
Toshio Uchida. 
Theory of force-free electromagnetic fields. II. Configuration with symmetry. 
Phys. Rev. E 56, 2198, 1997.

\bibitem{Stern70}
David Stern.
Euler Potentials.
American Journal of Physics 38, 494 (1970); doi: 10.1119/1.1976373, 1970.

\bibitem{Brandenburg11} 
Axel Brandenburg.
Magnetic field evolution in simulations with Euler potentials.
Mon. Not. R. Astron. Soc. 401, 347–354.
doi:10.1111/j.1365-2966.2009.15640.x, 2010.

\bibitem{Zaharia08}
Sorin Zaharia.
Improved Euler potential method for three-dimensional magnetospheric equilibrium.
Journal of Geophysical Research, Vol. 113, A08221, doi:10.1029/2008JA013325, 2008.

\bibitem{Komissarov12}
S. S. Komissarov. 
Time-dependent, force-free, degenerate electrodynamics. 
Monthly Notices of the Royal Astronomical Society, 
336(3):759–766, 2002.

\bibitem{Pfeiffer13}
Harald P. Pfeiffer and Andrew I. MacFadyen. 
Hyperbolicity of Force-Free Electrodynamics.
arXiv:1307.7782 [gr-qc], 2013.

\bibitem{Carrasco16}
Federico Carrasco and Oscar Reula.
Covariant Hyperbolization of Force-free Electrodynamics.
Phys. Rev. D, 93.93.085013, 2016.

\bibitem{Abalos15}
Fernando Abalos, Federico Carrasco, Erico Goulart, and Oscar Reula. 
Nonlinear electrodynamics as a symmetric hyperbolic system. 
Phys. Rev. D, 92:084024, 2015.

\bibitem{Kreiss89}
Heinz-Otto Kreiss and Jens Lorenz.
Initial Boundary Value Problems and the Navier-Stokes Equations.
Academic Press, San Diego, USA,
ISBN 0-12-426125-6, 1989.

\bibitem{Penrose84}
Roger Penrose and Wolfgang Rindler.
Spinors and Spacetime, Volume 1: two spinors calculus and relativistic fields.
Cambridge University Press.
ISBN 0 521 24527 3, 1984.

\bibitem{Carter79}
Brandon Carter, 
in General Relativity: An Einstein Centenary Survey,
edited by S. W. Hawking and W. Israel Cambridge University
Press, Cambridge, England, 1979.

\bibitem{Kreiss-Ortiz01}
Keinz-Otto Kreiss and Omar E. Ortiz.
Some Mathematical And Numerical Questions Connected With First And 
Second Order Time Dependent Systems Of Partial Differential Equations.
arXiv:gr-qc/0106085.
Lect.Notes Phys. 604 (2002) 359.
2001.

\bibitem{Reula98}
Oscar Reula.
Hyperbolic Methods for Einstein's Equations.
Living Reviews in Relativity 
Electronic Publication www.livingreviews.org/Articles/Volume1/1998-3reula/, 1998.

\bibitem{Reula04}
Oscar Reula.
Strongly hyperbolic systems in General Relativity.
Journal of Hyperbolic Equations, Vol. 1, No. 2 (2004) p. 251-269.
2004.

\bibitem{Geroch96}
Robert P. Geroch. 
Partial differential equations of physics.
[Online Los Alamos Archive Preprint]: cited on 19 January 1998, http:
//xxx.lanl.gov/abs/gr-qc/9602055. Scottish Summer School in Theo-
retical Physics. 1996.

\bibitem{Kreiss70}
Heinz-Otto Kreiss. 
Initial boundary value problems for hyperbolic systems. 
Commun. Pur. Appl. Math., 23:277–298, 1970.

\bibitem{Taylor91}
Michael E. Taylor.
Pseudodifferential operators and nonlinear PDE.
Progress in Mathematics 100, Birkhäuser, Boston, USA, 1991.

\bibitem{Strang66}
Gilbert Strang. 
Necessary and insufficient conditions for well-posed Cachy problems.
J.Differential equations, 2 p.107-114, 1966.

\bibitem{Friedrichs54}
Kurt Otto Friedrichs. 
Symmetric hyperbolic linear differential equations.
Comm. Pure Appl. Math. 7 (1954), 345-392.
1954.

\bibitem{Thorne82}
Kip S. Thorne and Douglas Macdonald. 
Electrodynamics in curved spacetime: 3+ 1 formulation. 
Monthly Notices of the Royal Astronomical Society, 198(2):339–343, 1982.

\bibitem{Smoller83}
Joel Smoller. Shock Waves and Reaction—Diffusion Equations. Grundlehren der mathematischen Wissenschaften 258 series, 2nd. edition, Springer US. 1983.

\bibitem{Leray53}
Jean Leray.
Hyperbolic differential equations, mimeographed notes.
Institute for Advanced Study, Princeton, 1953.

\bibitem{Leray-Ohya67}
Jean Leray and Yujiro Ohya.
Équations et systèmes non-linéaires, hyperboliques nonstricts.
Math. Ann. 170, pp. 167–205, 1967. 

\bibitem{Choquet82}
Yvonne Choquet-Bruhat, Cecile Dewitt-Morette and Margaret Dillard-Bleick.
Analysis, manifolds and physics.
Vol. 1, Revised Edition, Elsevier, 1982.

\bibitem{Choquet08}
Yvonne Choquet-Bruhat.
General Relativity and the Einstein Equations.
Oxford Mathematical Monographs, 0199230722, 2008.

%\bibitem{Perlick11}
%Volker Perlick.
%On the hyperbolicity of Maxwell’s equations with a local constitutive law.
%Journal of Mathematical Physics, 52(4):042903. 2011.

\bibitem{Wald84}
Robert Wald.
General Relativity.
University of Chicago Press.
ISBN-10: 0226870332, 1984.


\end{thebibliography}

\end{document}